\documentclass{cplslarge}%
\usepackage[authoryear]{natbib}
\usepackage{graphics, epstopdf}
\usepackage{makeidx}
\makeindex

\begin{document}
\author{B.T. Nadiga and D.N. Straub}
\chapter{Meridional Propagation of Zonal Jets in Ocean Gyres}

\section{Introduction}
Analyses of both altimetric data and in-situ measurements reveal
patterns of meridionally-alternating, nearly zonal, coherent jet-like
structures in many of the world ocean basins
\citep{maximenko2005observational, maximenko2008stationary,
  ivanov2009system, melnichenko2010quasi, van2011quasi}.  A number of
realistic Ocean General Circulation Model (OGCM) simulations also
display similar structures \citep[e.g.][]{treguier2003origin,
  nakano2005series, richards2006zonal}.  Most such analyses, whether
model- or altimetry-based, consider averages, typically over several
months, of geostrophic velocity.  Generally, these jet-like structures
i) extend zonally from just a few degrees to tens of degrees of
longitude, ii) display meridional wavelengths in the range of one to
as many as five degrees of latitude, iii) show a high degree of
vertical coherence, and iv) have speeds of a few centimeters per
second.  In many instances, the jet orientation is slightly off-zonal.
For brevity we will refer to these structures as alternating zonal
jets, or simply, jets.

It has been suggested
that the jets may be simple artifacts of averaging over
westward propagating eddies \citep[e.g.,][]{schlax2008influence}.
More commonly, the jets are assumed not to be artifacts,
and many dynamical formation mechanisms have been put forth.  For example,
alternating jets are well-known from $\beta$-plane turbulence and are
associated with a halting of the two-dimensional inverse energy
cascade by Rossby wave dispersion \citep[e.g.,][]{rhines1975waves,
  vallis1993generation}. It is thus possible that the observed
alternating jets in the oceans are a result of the anisotropic inverse
cascade mechanism \citep{kramer2006beta, nadiga2006zonal}.  They may
also result from nonlinear self-interactions of linear eigenmodes
\citep{berloff2009mechanism} or from from radiating instabilities of
unstable eastern boundary currents \citep{hristova2008radiating,
  wang2012new}.  They may be related to preferred growing structures
excited by the imposed stochastic forcing \citep{farrell2007structure,
  bakas2013mechanism} or be manifestations of the $\beta$-plume effect
\citep[e.g.][]{afanasyev2012origin}.  Finally, it has been suggested
that the jets might be formed by an instability of Rossby waves
\citep{lorenz1972barotropic, gill1974stability,
  connaughton2010modulational}.  As with altimetry-based and realistic
OGCM-based analyses of the jets, most of the process studies that
motivate these various formation mechanisms also consider temporal
averages of geostrophic velocity anomalies. Additionally, issues
related to meridional propagation of the jets is generally not
emphasized, and is explicitly treated only in the last set of studies (which
considers the formation of zonal and off-zonal jet structures in the
context of Rossby wave instability).  It is also only in this set of
studies that issues of off-zonality are explicitly considered.

The recent analysis by \cite{penduff_anim}
of one of his OGCM simulations is one case in which off-zonal
orientation and meridional propagation are seen. In one of his
climatology-forced ocean/sea-ice simulations---ORCA025-MJM01, an
animation of the 18 month-running average of SSH over a dozen years
makes evident the propagating nature of alternating jets that form in
his simulations \citep{penduff_anim}.  In recent quasigeostrophic (QG)
simulations of wind driven gyres, we have found jets be
near-ubiquitous and to propagate in a manner similar to Penduff's
animation.  In the QG simulations, both the jets and their propagation
are clearly visible in the instantaneous fields ---without the need
for time averaging. They are especially clear in snapshots of the
baroclinic zonal velocity, for example.  Given that the propagating
nature of the jets and their presence in instantaneous snapshots of
model simulations has not yet received much attention, we focus on
these aspects of alternating jets for most of this chapter.  We will
also comment on the formation mechanism in our two layer
quasigeostrophic $\beta$-plane simulations.  Specifically, we will
consider whether the jets might be formed via an instability 
Rossby waves, which are also ubiquitous in our simulations.

\cite{connaughton2010modulational} consider modulational instability
of Rossby waves in the barotropic and equivalent-barotropic settings,
extending the analysis of \cite{gill1974stability}.  For weak waves
($M = \psi_o p^3/\beta \ll 1$ where $\psi_o$ is the initial amplitude
of the Rossby wave of wavenumber $p$ and $\beta$ is the usual
meridional gradient of Coriolis frequency), they show that maximum
growth occurs for off-zonal modulations that are close to being in
three-wave resonance with the primary wave.  For strong waves ($M \gg
1$), the most unstable modes are perpendicular to the primary wave.
For a strong primary wave having a zonally-oriented wavevector, the
growing modes thus have meridionally-oriented wavectors ($k_x = 0$),
corresponding to zonal flow.  Note also that the Rossby wave dispersion
relation ($\omega = -\beta k_x/ (k^2 + L_d^{-2})$ ---where $\omega$ is
the wave frequency, $k_x$ is the zonal wavenumber, $k$ is the total
horizontal wavenumber and $L_d$ is the Rossby deformation radius--- then
has that jets excited by this mechanism would be steady.  On the other
hand, when the primary Rossby wave is weak, off-zonality of the
unstable mode leads to propagation of the off-zonal jets, again due to
nature of the dispersion relation.

\begin{figure*}
\figurebox{40pc}{50pc}{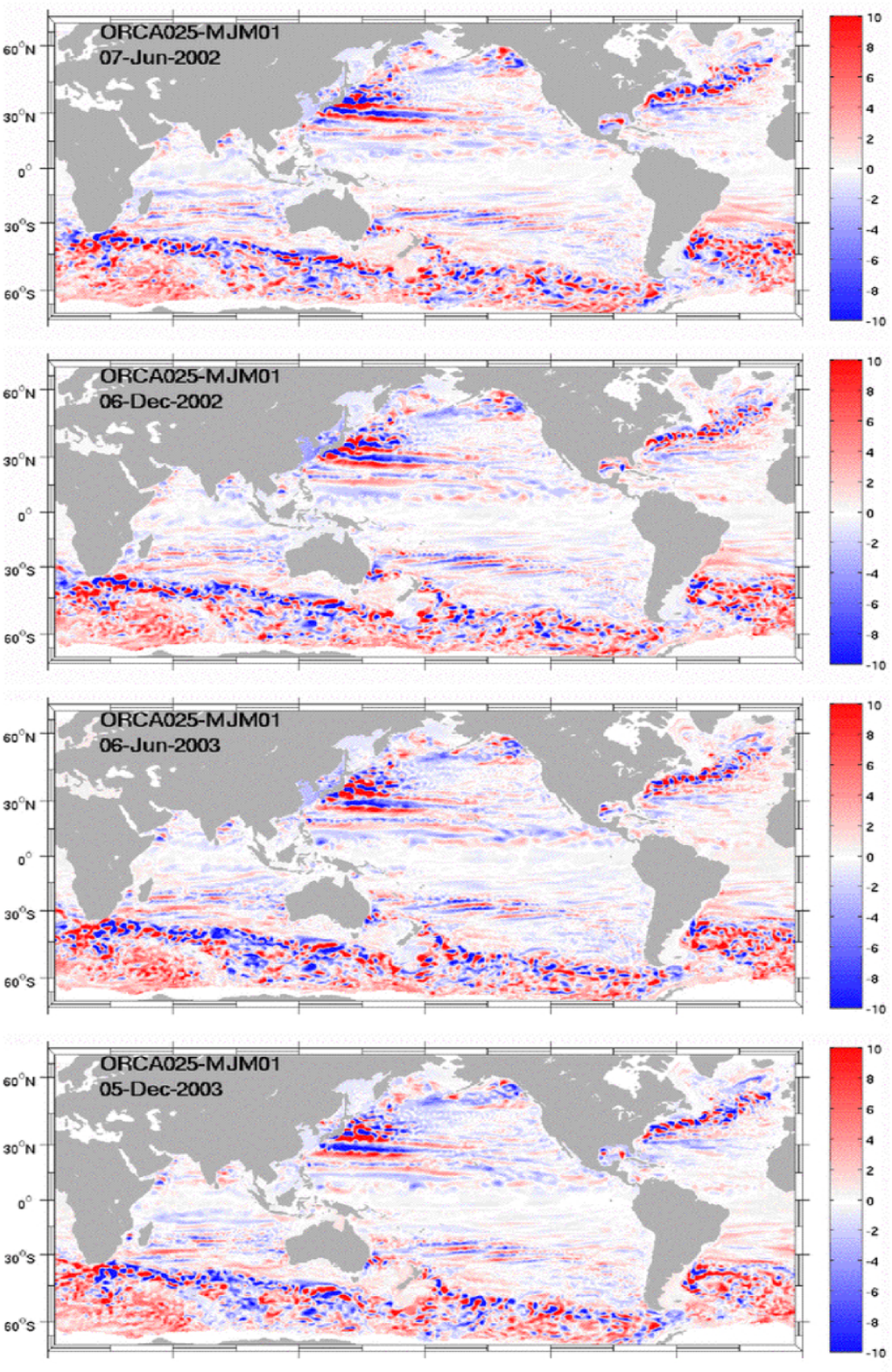}
\caption{Snapshots from the \cite{penduff_anim} animation mentioned in the text. 
Time increases from top to bottom in 26 week intervals; 
focusing attention on the meridional position
of the jets in successive snapshots shows a clear meridional propagation.}
\label{anim}
\end{figure*}
\clearpage

\begin{figure*}
\figurebox{30pc}{}{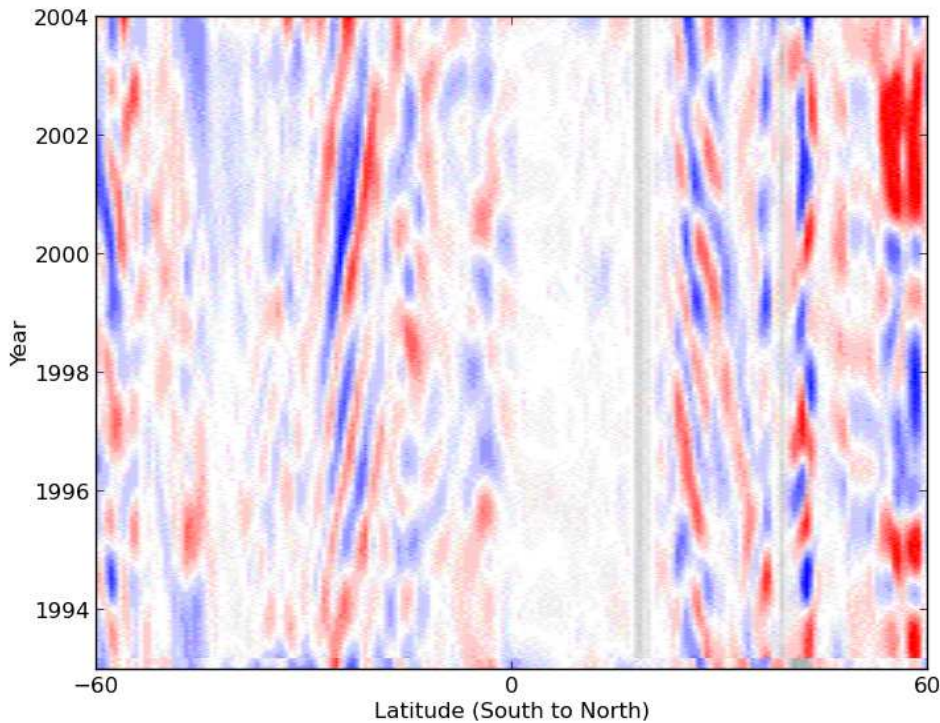}
\caption{Hovmoller plots in the western Pacific from Penduff's animation.  
Equatorward propagation of jets in sub-tropical gyres (around
  30 degrees North and South) is evident; the rate of propagation is
  roughly about one degree of latitude per year. There are also hints of
  poleward propagation of the jets in the North Pacific sub-polar gyre
 (around 60 degrees North).}
\label{hov_anim}
\end{figure*}

The Rossby wave instability mechanism discussed above is independent
of the source of the primary Rossby waves, which could be triggered by
any number of processes. For example, in the oceanic context, long baroclinic Rossby
waves
can be radiated from the eastern boundary, either due to unstable
boundary currents there \citep[as, e.g., in ][]{hristova2008radiating,
  wang2012new} or due to energy arriving there in the form of Kelvin
waves. 
In a multi-layer QG
basin, mass conservation leads to a parameterization of Kelvin waves
that determines the value of baroclinic streamfunctions on the basin
perimeter.  This can lead to oscillations in the value of the streamfunction
(equivalently, the thermocline height) along the boundary,
and these oscillations can  excite Rossby waves.  In particular, long
Rossby waves generated in this way are commonly seen propagating
westward from the eastern boundary in quasigeostrophic gyre
simulations.  In our QG simulations, we also see barotropic Rossby
waves, as well as a band of  Rossby-wave-like
features that  have a mixed baroclinic-barotropic character and 
lie adjacent to the region populated by jets.

\cite{nadiga2010alternating} considered the barotropic $\beta-$plane
vorticity equation in a closed basin with steady double gyre forcing
and found the spontaneous appearance of alternating jets in long time
averages.  The jets were associated with a weakly-forced,
weakly-dissipated limit, and also evident in this limit was a ``double
cascade" of energy. That is, the upscale nonlinear transfer (familiar
from classic studies of 2-dimensional turbulence) was offset by a
forward transfer due to the $\beta$ term. That the linear $\beta$ term
could transfer energy at all is related to the basin geometry; in a
periodic setting, this term is not directly associated with energy
transfer between scales.  \cite{straub2014energy} considered the
two-layer double gyre problem and also found both quasi-zonal jets and
a double cascade of barotropic energy.  They did not emphasize the
jets or find any clear link between their characteristics and the
double cascade.  They did, however, find the jets to be present over a
wide range of parameters and to be easily visible in snapshots of the
baroclinic streamfunction.  Aspects of the jets in this baroclinic QG
double gyre setting will be discussed in Section 3.

\adjustfigure{70pt}
\section{Alternating jets in a  $\frac{1}{4}^o$ OGCM Simulation}

\begin{figure*}
\figurebox{30pc}{}{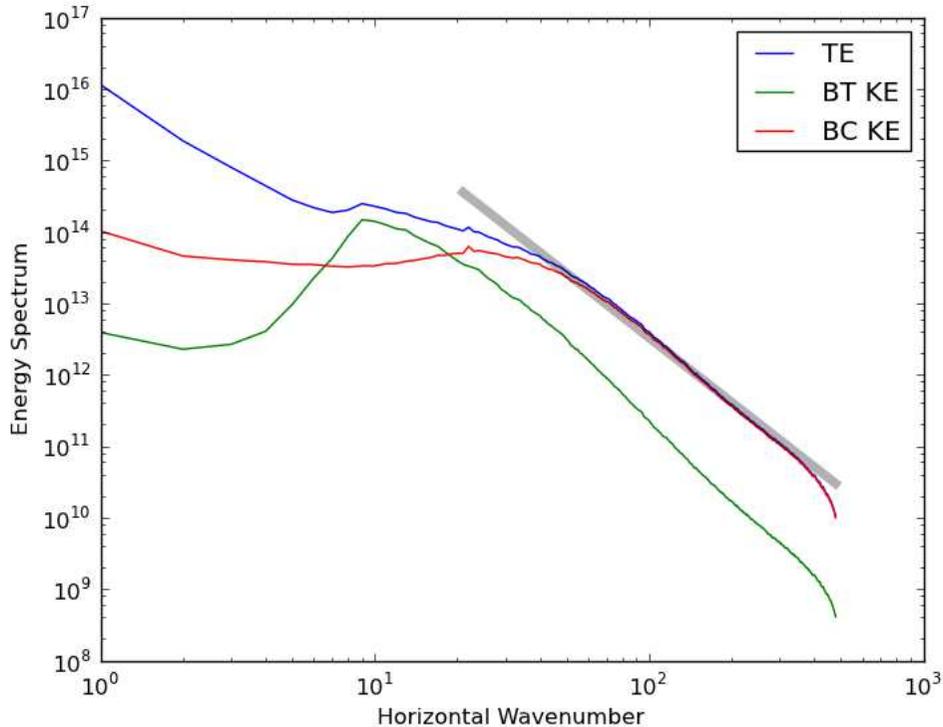}
\caption{Time-averaged spectra of barotropic and baroclinic kinetic
  energy and of total energy for our reference QG simulation.}
\label{lr02_spec}
\end{figure*}

\cite{penduff2010impact} considered global ocean/sea-ice simulations
driven by realistic atmospheric forcing over the period 1993-2004. 
They considered four horizontal resolutions  (2$^o$, 1$^o$, $\frac{1}{2}^o$, and
$\frac{1}{4}^o$) and 
compared sea level
anomalies (SLA) against the AVISO SLA dataset. In various measures,
they found a monotonic improvement in the comparison with
resolution. In conjunction with their $\frac{1}{4}^o$ simulation, they
also conducted a companion simulation---ORCA025-MJM01. The only
difference with this simulation was that it was forced by an
annually-repeating average of the atmospheric forcing fields
\cite{penduff_anim}. The reader is referred to \cite{penduff_anim} for
an animation of the 18-month running average of SSH in this companion run.
In this animation, the propagating nature of the
alternating jets is evident in the subtropical and subpolar gyres of
the North Pacific and in the supolar gyre of the South Pacific. For
reference, Fig.~\ref{anim} shows four snapshots from that animation of
sea surface height (SSH) separated by 26 weeks each. Not only is the
widespread appearance of alternating banded structures evident in
these snapshots, but the propagation of these structures may be
surmised by focusing attention on their meridional location 
in the successive snapshots going from top to bottom.

To better demonstrate this propagation, Fig.~\ref{hov_anim} shows the
time evolution of SSH as a Hovmoller diagram. In this diagram, the
equatorward propagation of the alternating jets in the northern and
southern subtropical gyres (around $\pm 30^o$) is evident. There are
also further hints of poleward propagation of similar structures in
the northern subpolar gyre, with the Antarctic Circumpolar Current
complicating the picture in the south. In the subpolar gyres, the
alternating jets are seen to propagate at speeds of roughly one degree of
latitude per year.

\section{Jets in the QG double gyre problem}

We have found alternating zonal jets to be a near-ubiquitous feature
of two layer QG double gyre simulations over a wide range of
deformation radii and forcing and dissipation parameters.  The jets
are easily visualized as quasi-zonal structures in the instantaneous
zonal velocity associated with the baroclinic mode. They are also
evident in snapshots of the barotropic zonal velocity, although in
this case larger scale structures are superposed on the jets.  The
character of the jets varies somewhat with parameters; it is
particularly sensitive to the deformation radius, for example.
Simulations assuming a large deformation radius show the central jet
stemming from the confluence of the two western boundary currents to
be relatively strong, with the alternating jets being fewer and less
pronounced.  For small deformation radii, the converse is true, and
the alternating jets are particularly robust.  Forcing and dissipation
strengths also affect the jets.  The jet region is bounded to the
east, north, and south by a band wavelike features, and this band is
more pronounced in weakly forced or strongly damped (bottom friction)
simulations.  We will consider later whether this band of waves might
play a role in establishing the jets or whether it might instead be
thought of as modified Rossby waves that propagate away from the jet
region.  Since the jets are particularly evident in low deformation
radius simulations, we will first consider this case ($L_d = 15$ km; 
referred to as our reference case). We then present results from
simulations a) that use a larger deformation radius ($L_d = 30$ km)
and b) that use a weaker wind forcing as compared to the reference
run.

\adjustfigure{20pt}
We emphasize that the jets appearing in this two-layer setting owe
their existence to an interaction involving both layers; equivalently,
involving both the barotropic and baroclinic modes. As mentioned
above, in other work we have also found alternating jets in barotropic
double gyre simulations; however, in the single layer setting, the
jets are weak and are visible only in long time averages.  Similarly, a
reduced gravity version of the barotropic equations does not produce
jets that are visible in snapshots.  It thus seems that the jets in a two-layer
setting are somehow fundamentally related to an interaction between
the barotropic and baroclinic modes.

\begin{figure*}
\figurebox{40pc}{}{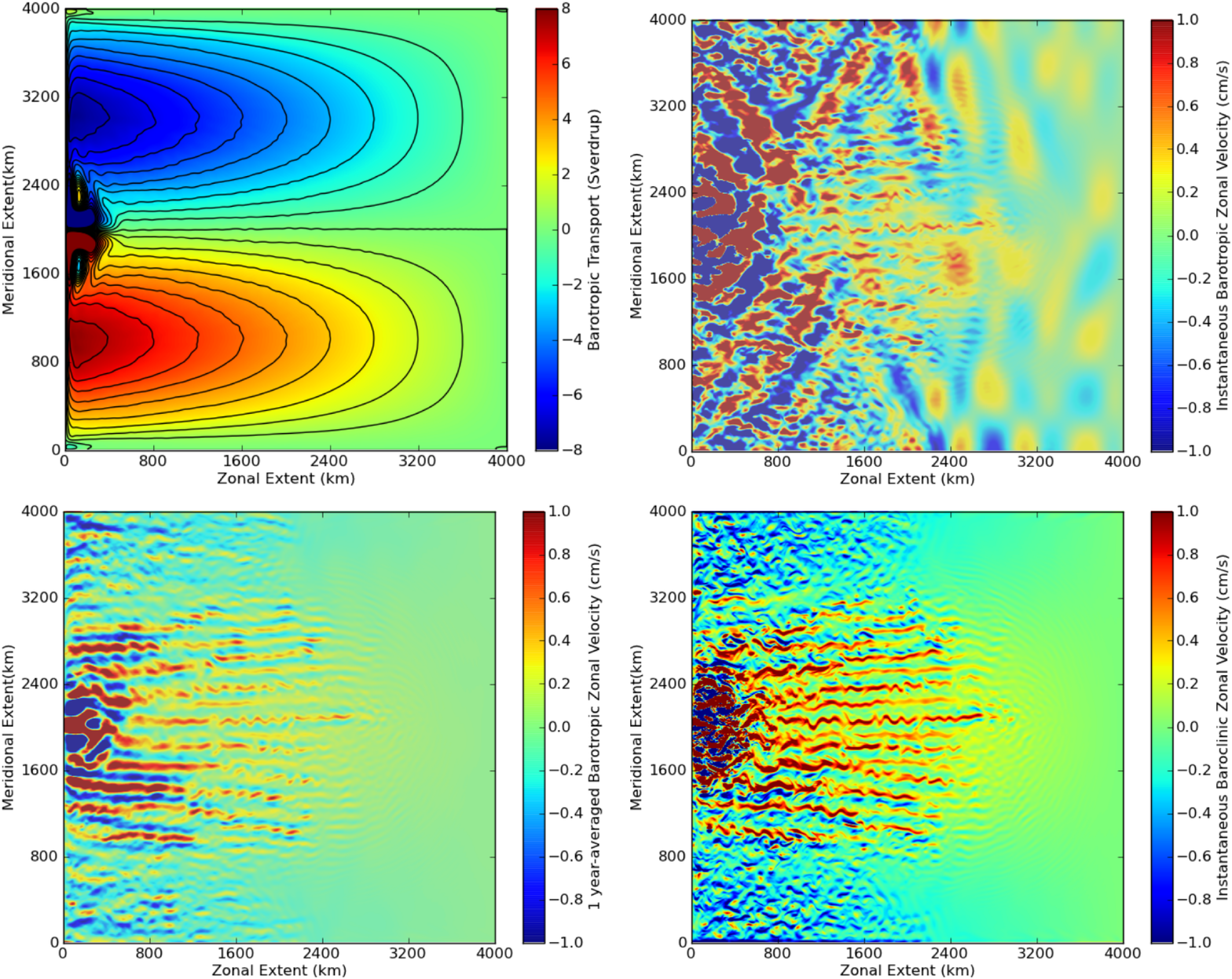}
\caption{Physical space fields for our reference simulation: Sixty year
  time average of barotropic streamfunction (upper left),
  instantaneous snapshot of the barotropic zonal velocity (upper
  right), one year average of the barotropic zonal velocity (lower
  left), instantaneous snapshot of the baroclinic zonal velocity
  (lower right).  }
\label{lr02_4}
\end{figure*}

We consider the quasigeostrophic equations truncated to two layers in
the vertical. 
Written in the layer form, the governing equations are 
\begin{eqnarray}
  {\partial q_i \over \partial t} + J({\psi_i}, q_i) + \beta v_i& = & \delta_{i1}F -
  \delta_{i2}r\nabla^2\psi_i + A\nabla^8\psi_i \nonumber \\ q & =& \nabla^2\psi_i
  + (-1)^{i}{f_0^2\over g'h_i}\left(\psi_1-\psi_2\right).\
\end{eqnarray} 
In the above equations, besides the standard notation used, the subscript
$i$ is the layer index: 1 for upper, 2 for lower, $q$ is quasigeostrophic potential
vorticity (QGPV), $\psi$ is streamfunction, $J(\psi_i, q_i) = 
\partial\psi_i/\partial x\,
\partial q_i/\partial y - 
\partial\psi_i/\partial y\,
\partial q_i/\partial x$ and represents horizontal advection of QGPV,
and $\delta_{ij}=1$ if $i=j$ and 0 otherwise (implying that the wind
forcing is applied only to the upper layer and bottom friction to the
lower layer). These equations are solved in a closed mid-latitude
$\beta$-plane domain that is discretized on a regular Cartesian
grid. For the biharmonic viscosity used, boundary conditions consisted
of setting both relative vorticity and its Laplacian to zero at the
horizontal boundary.  A nominally fifth order adaptive time-stepping
Runge-Kutta Cash-Karp method was used for time integration and a
direct solve was used for inverting the quasigeostrophic potential
vorticity-streamfunction equation.

\adjustfigure{55pt}

\begin{figure*}
\figurebox{40pc}{}{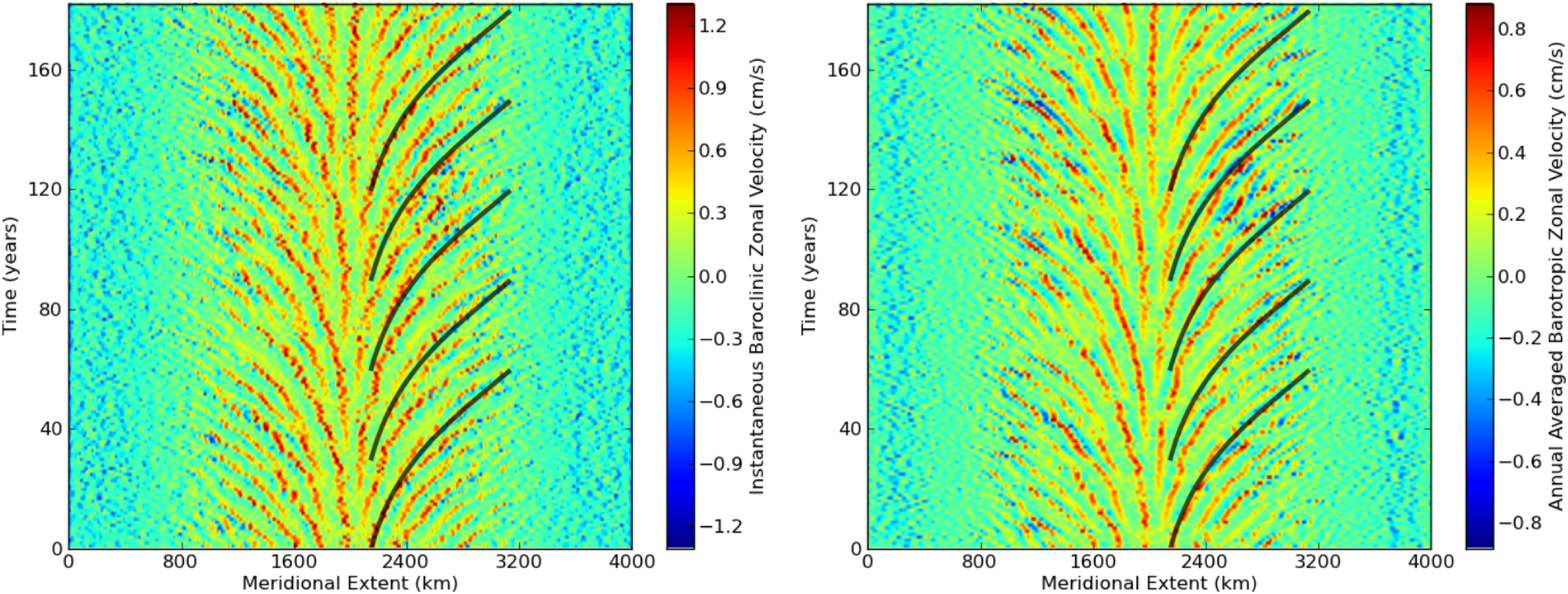}
\caption{Hovmoller plots of instantaneous baroclinic (left) and one
  year-averaged barotropic (right) zonal velocities for our reference
  simulation.  The zonal location considered is 1600 km from the
  western boundary. The jets are seen to slowly propagate equatorward in
  the sub-tropical gyre and poleward in the sub-polar gyre. Reference
  black curves drawn are propagated outwards from the central latitude
  by the time-mean baroclinic meridional velocity. }
\label{lr02_hov}
\end{figure*}

We revisit the classic baroclinic wind-driven double gyre problem.
The domain consists of a 4000 km square $\beta$-plane basin with upper
and lower layer thicknesses of 1 km and 4 km, respectively. The
horizontal domain is spanned by 960 points in each direction to yield
a horizontal grid spacing $\Delta x \approx 4.2 \mbox{km}$. The
reference simulation uses a first internal Rossby deformation radius
of 15 km.  A steady sinusoidal double gyre windstress with a maximum
amplitude of 0.1 Nm$^{-2}$ is applied to the upper layer.  The
reference drag coefficient is set to $10.67 \times
10^{-8}\mbox{s}^{-1}$.  A biharmonic form of horizontal dissipation
was used to dissipate the forward-cascading potential enstrophy; the
coefficient of this biharmonic viscosity was chosen to scale as $A =
\beta \Delta x^5$, with $\beta = 2 \times
10^{-11}\mbox{m}^{-1}\mbox{s}^{-1}$.  This choice allows for a smooth
tapering of the energy spectra at high wavenumber, and is sufficiently
small such that dissipation due to the biharmonic term is small.  The
one dimensional horizontal spectra of barotropic and baroclinic
kinetic energy, total energy, and a reference $-3$ slope are shown in
Fig.~\ref{lr02_spec}.

Figure~\ref{lr02_4} shows various physical space fields for our
reference simulation.  The long-time ($\approx$180 year) average
(upper left) of the barotropic streamfunction shows the double gyre
structure, with an approximate Sverdrup balance
\citep[e.g.][]{pedlosky87} over most of the domain.  Also evident,
near the western boundary, are tight inertial recirculations. These
are confined to a region close to the boundary; this apparently being
a consequence of the relatively small deformation radius (for larger
deformation radii, these recirculations extend farther east; not
shown).  There is no clear evidence of jets in the long-time average.
By contrast, the instantaneous barotropic velocity (upper right) does
show evidence of jets.  Superposed on the jets are larger scale
diagonally-oriented features that appear to be barotropic Rossby
waves;
they do not appear in the baroclinic mode and 
animations show them to propagate
rapidly with westward phase velocities.  Similar wave structures also
obscure the jets in the upper layer velocity field.  Time averaging
serves to remove these structures (lower left panel), so that the jets
can be more easily visualized.  We emphasize that time averaging makes
the jets visible by removing structures superposed on the jets; it is
not the case that the jets result from time averaging of propagating
eddies.  Put another way, the jets can also be made more evident by a
spatial filtering of snapshots, without the need for time averaging
(not shown).  By contrast, no such filtering (in space or time) is
needed to make the jets evident in the baroclinic mode. Instead, in  the baroclinic mode,
the jets are clearly visible in unprocessed snapshots (lower right
panel).

From visual inspection, it appears that the barotropic and baroclinic
jets evident in snapshots are related.  Moreover, animations show the
jets associated with each mode to propagate outward from the center
latitude at comparable speeds.  This propagation is also evident in
Fig.~\ref{lr02_hov}, which shows Hovmoller plots of two of the fields
depicted in Fig.~\ref{lr02_4}.  Jets appear to form near the center
latitude and propagate meridionally outwards. Note that the
propagation speed is latitude-dependent, being fastest near the
(latitudinal) centers of the two gyres (e.g., near 1000 or 3000 km on
the horizontal axes in Fig.~\ref{lr02_hov}).  This suggests that the
propagation speed might scale with the meridional component of the
Sverdrup velocity, which also peaks at these same latitudes.  The
curves overlain in Fig~\ref{lr02_hov} emphasize this point.  Plotted
is a family of curves, $y(t)$. Each curve begins at a small distance,
$y_0$, north of the center latitude and obeys $dy/dt = v_{BC}(y)$,
where $v_{BC}$ is the time averaged baroclinic meridional velocity, or
about twice the depth-averaged meridional velocity.\footnote{The
  linear modes are normalized such that $\psi_{barotropic} =
  \epsilon_1 \psi_1 + \epsilon_2 \psi_2$ and $\psi_{baroclinic} =
  \epsilon_3(\psi_1 - \psi_2)$, where $\epsilon_1$ and $\epsilon_2$
  are the fractional layer thicknesses (here 0.2 and 0.8,
  respectively), and $\epsilon_3 = (\epsilon_1\epsilon_2)^{1/2}$. }
The curves fit reasonably well with the observed propagation. Exactly
why this should be the case remains unclear.  The figure does provide
clear evidence, though, that the meridional propagation of the jets is
in large part due to advection by the gyres.

\begin{figure*}
\figurebox{40pc}{}{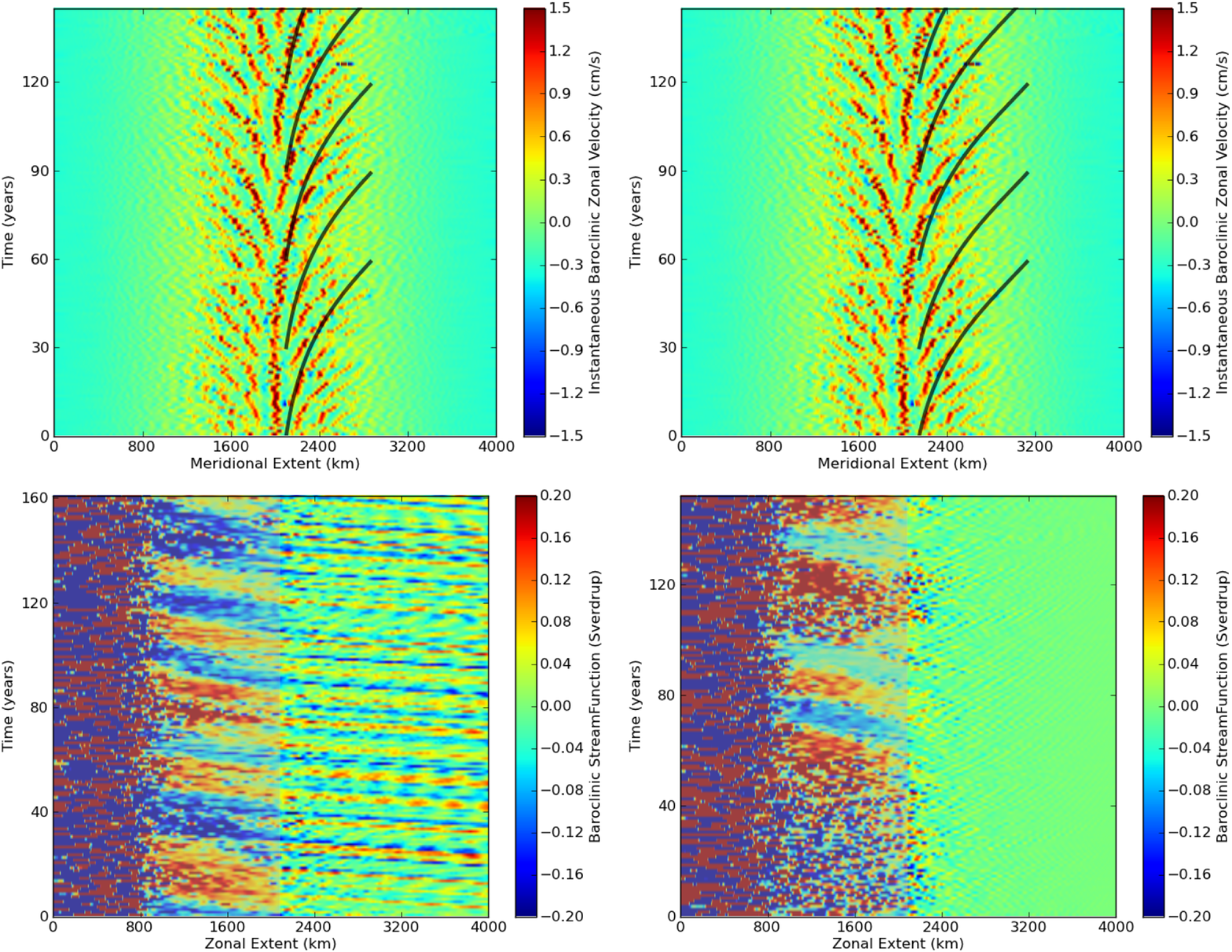}
\caption{
Hovmoller diagrams for our 30 km deformation radius
  simulation.  Upper panels are as in the left panel of 
  Fig.~\ref{lr02_hov}. 
  Lower panels plot the
  instantaneous baroclinic streamfunction at $y = 2000$ km and show
  zonal propagation.  Left
  panels keep the wall mode ($\tau$) and right panels do not (i.e.,
  the right panels are for a companion simulation in which $\tau$ was
  artificially set to zero at each time step).  A ``waterstain" has been
  added to the left side of the lower panels for added visual clarity.  
 }
\label{lr04}
\end{figure*}

The top left panel of Fig.~\ref{lr04} is analogous to the left panel
in Fig.~\ref{lr02_hov}, but for a more realistic value of the
deformation radius ($L_d = 30$ km; all other parameters are as in our
reference simulation).  As anticipated, jets are wider and fewer when
the deformation radius is increased.  Also as anticipated, meridional
propagation is also evident here.  The overlain curves are analogous
to those in Fig.~\ref{lr02_hov}. Although the fit is less impressive
than before, there nonetheless remains the suggestion that advection
by the gyres plays a key role in the meridional propagation.

\adjustfigure{65pt}

The bottom left panel in Fig.~\ref{lr04} is an $x$-$t$ Hovmoller plot
of the baroclinic streamfunction.
It shows a  broad pattern of westward propagation
extending back from  about $x = 2000$ km. This appears to be low frequency 
Rossby waves, possibly 
generated by the break-up of the midlatitude jet at about this same
longitude. 
Note also the clear signature of westward propagating
long Rossby waves stemming from the eastern boundary.  As
mentioned in the introduction, it is possible that the jets are formed
in association with an instability of these waves.  Indeed,
\cite{o2012emergence} found that addition of a fast time scale
stochastic single-gyre forcing to the classic double gyre problem
resulted in robust long Rossby waves; and related to these, jets.  In
our simulations, the forcing is steady and the Rossby waves are
instead formed in association with a lifting and lowering of the wall
value for the baroclinic streamfunction.  That is, at each time step a
function, $\tau(x,y)$, must be added to the
potential-vorticity-containing part of the baroclinic streamfunction.
The function $\tau(x,y)$ obeys
\begin{eqnarray}
\nabla^2\tau - \frac{1}{L_d^2}\tau = 0
\end{eqnarray}
\noindent
with $\tau = \tau_0$ on the boundary.  Here, $\tau_0$ is chosen such
that the area integral of $\tau$ is equal and opposite to that of the
potential-vorticity-containing part of the baroclinic streamfunction.
The spatial structure of $\tau$ is that it decays over a distance
$L_d$ from its value on the perimeter to a value of zero in the
interior.  Typically $\tau_0$ varies with time, and this lifting and
lowering of the thermocline excites Rossby waves.

To test whether an instability of these long Rossby waves might be
central to the formation of the jets, we ran a companion simulation in
which this zero potential vorticity mode was artificially suppressed
(i.e., $\tau_0$ was set to zero).  The right panels of Fig.~\ref{lr04}
are analogous to the left ones, but for the companion simulation.
Clearly, this suppression filters the long Rossby waves emanating
from the eastern boundary, but does not filter the jets.  Both the
jets and their meridional propagation persist in the absence of these
long Rossby waves; in fact, the two upper panels suggest that the jets
are virtually unaltered when the wall mode is suppressed.  This was
also the case in our 15 km simulation; although there, the long Rossby
waves (present when the zero potential vorticity was retained) were
weaker and associated with a lower band of frequencies. Additionally,
snapshots of the baroclinic zonal velocity were virtually
indistinguishable between cases where the zero potential vorticity mode
was kept and those where it was not.  We thus conclude that the jets
are not formed in association with instability of Rossby waves
generated at the eastern boundary.

\adjustfigure{70pt}
\begin{figure*}
\figurebox{40pc}{}{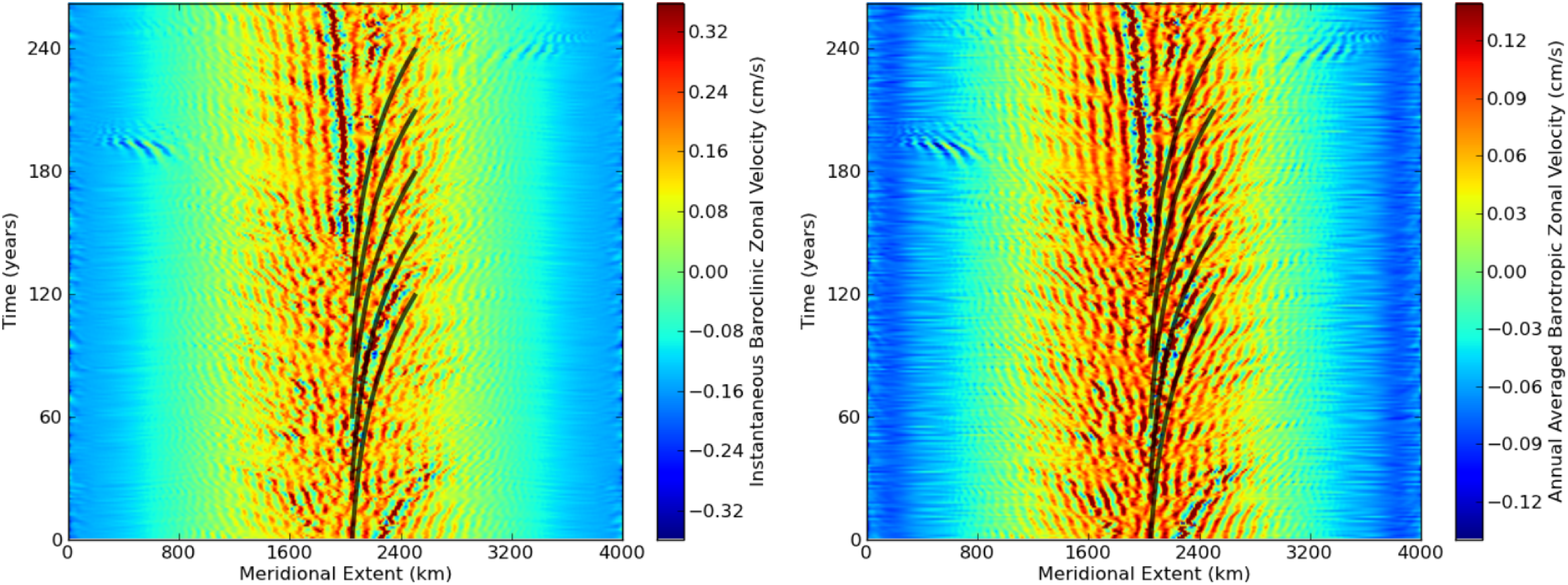}
\caption{Similar to Fig.\ref{lr02_hov}, but for our weakly forced simulation.  
The alternating zonal jets are seen to get weaker with
respect to the gyre circulation when the windstress reduced by a
factor of 2. The meridional propagation also slows down, suggesting
  that it is related to advection by 
  the Sverdrup velocity.}
\label{lr02_weak}
\end{figure*}

\begin{figure*}
\figurebox{40pc}{}{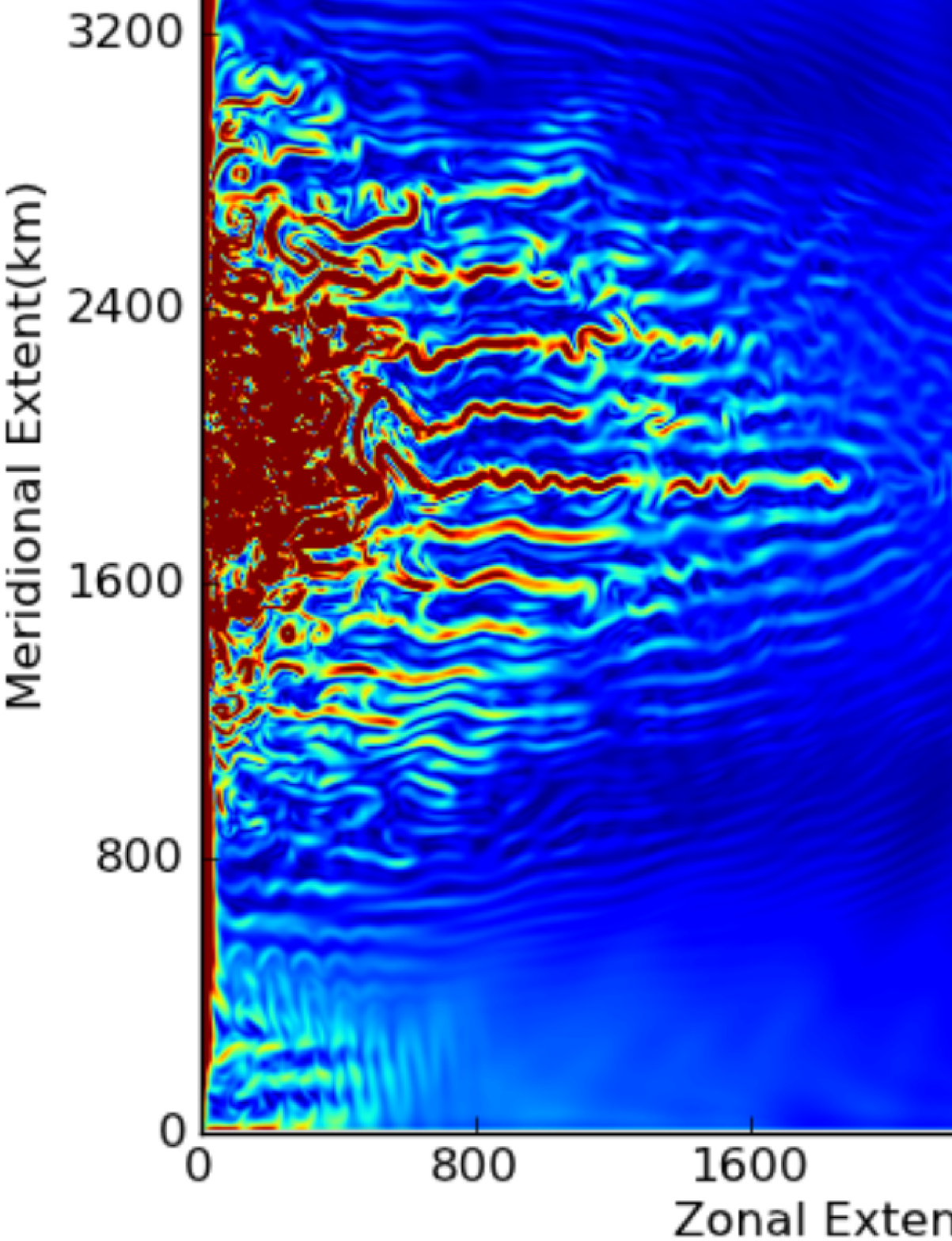}
\caption{Baroclinic speed for our weakly forced simulation.
Values have been capped so as to emphasize
the band of short Rossby waves surrounding the jet region. }
\label{ref_radiation}
\end{figure*}

Figure ~\ref{lr02_weak} shows Hovmoller diagrams analogous to those in
Fig.~\ref{lr02_hov}, but for a simulation forced by weak winds
(maximum stress of 0.05 Nm$^{-2}$ as opposed to 0.1 Nm$^{-2}$ for our
reference simulation). As before, the plots are scaled so as to
emphasize the jets.  This scaling makes not only the jets, but also
the zonal velocity associated with the gyres evident.  Such was not
the the case in Fig.~\ref{lr02_hov}.  In other words, relative to the
gyres, the jets are weaker here.  Also as before, curves indicative of
advection by the baroclinic Sverdrup velocity are overlain.  The fit
of these curves with the jets near the center latitude is quite good;
however, it is less good near the center of the gyres, where $v_{BC}$
is maximal and, at times, the jets appear to stall.  Further out form
the center (e.g., around $y = 3000$ km in the north or $y = 1000$ km
to in the south) lies a band of wavelike structures.  This can be seen
by enlarging Fig.~\ref{lr02_weak}, but is blurred in the print
version.  Wave bursts sometimes appear (e.g., around days 180 and 240
in Fig.~\ref{lr02_weak}) and produce packets that propagate outward
from the jet region.
More typically, these wavelike features remain adjacent to the jet region.  A
physical space picture is given in 
Fig.~\ref{ref_radiation}. Plotted
is the instantaneous baroclinic speed for our weak wind simulation,
scaled so as to make this relatively weak band
of wavelike structures more evident. 
A similar waveband is also seen in the barotropic mode 
and an outward group velocity can be inferred from a
Hovmoller diagram (not shown); the outward group velocity suggest that it
is more natural to think of these features as excited by the jet-turbulence region,
rather than as playing a significant role in forming the  jets.

In the introduction, we mentioned the possibility that the jets may be
formed by an instability of Rossby waves.   
We have seen four types of waves or wavelike features evident in our simulations.
These are a) the band of mixed barotropic-baroclinic waves discussed above, b) the
long baroclinic Rossby waves stemming from the eastern boundary, c) 
the barotropic Rossby waves (e.g., evident in the upper right panel of Fig ~\ref{lr02_4}) and  
d) the broad baroclinic westward propagation extending back from near the eastern
edge of the jet region (e.g., lower panels of Fig ~\ref{lr04}).  We have also presented
evidence that the first two are not likely responsible for forming the jets. The barotropic
waves, present whether or not the wall mode is kept) may be involved in an instability 
leading to the jets; unfortunately, we had no simple way to filter them, and so were not
able to test this numerically.  The broad westward propagation back from near the center
longitude of the basin (d, above) is the least well-characterized of these four types. 
These features appear to form near the eastern extent of the jet region and propagate
back towards the west. This suggests that their dynamics are somehow linked to those of the
jets.  On one hand, they appear to be formed by the jets; on the other, they propagate
back into the jet region, and hence may be involved in sustaining the jets, once formed.

\adjustfigure{60pt}

\section{Conclusions}

In this chapter, we have emphasized the meridional propagation of jets
embedded in mid-latitude ocean gyres.  We also emphasized that the
jets are clearly evident in snapshots, particularly of the baroclinic
mode. The meridional propagation seems related to an advective
mechanism. In particular, the propagation speed appears to agree well
with the baroclinic gyre velocity. The agreement is not perfect.
Also, why the advecting velocity should be $v_{BC}$ (equivalently,
twice the depth-averaged Sverdrup velocity) is unclear.  What forms
the jets also remains unclear. Jets evident in snapshots do not appear
in barotropic or single layer reduced gravity simulations, suggesting
that their formation is somehow fundamentally related to an
interaction between the barotropic and baroclinic modes. In the
introduction, we mentioned formation mechanisms involving
instabilities of Rossby waves.  In simulations not presented here, we
have also seen jets to be formed by similar instability
mechanisms. For example, artificially lifting and lowering the
thermocline height along the boundary (i.e., $\tau_0$) produces Rossby
waves that propagate inward from the basin boundaries.  Simulations
driven solely by this mechanism (i.e., with no potential vorticity
forcing at all) can produce wave trains that go unstable to form zonal
jets.  This occurs, however, only when the forcing is sufficiently
strong.  In our wind-driven gyre simulations, the Rossby waves
generated naturally by this dynamics are relatively weak, and this
does not appear to be the mechanism responsible for the jets seen in
these simulations.  For example, artificially suppressing the wall
mode produced jets that were virtually indistinguishable from those in
simulations where this wall mode was kept.  In other periodic
two-layer simulations (also not presented), we have seen short Rossby
waves go unstable to produce, at least temporarily, jet-like features.
Our gyre simulations do show the presence of short Rossby-wave-like
features surrounding the jet region.  These waves are of a mixed
barotropic-baroclinic nature and appear to be associated with eastward
group velocities.  As such, it seems more natural to interpret them as
produced by the turbulent dynamics of the boundary current confluence
region.  Consistent with this idea, these waves are energetically
weak.

Having ruled out a number of formation mechanisms, we note that a
hypothesis that we are currently investigating is that the jets result
from an instability of the main eastward flow associated with the
confluence of the subtropical and subpolar gyres.  Since individual
jets can typically be traced (e.g., using animations or Hovmoller
plots) back to near the center jet, it seems likely that the jets form
as an instability of this ``main jet".  In more realistic simulations
(such as the Penduff simulations highlighted in section 2), this would
correspond to the eastward extension of, say, the Gulf Stream or
Kuroshio.  It may be possible to understand the weaker, alternating,
jets seen on either side of these major currents as resulting from an
instability (perhaps involving waves propagating through the region)
of these major currents.

\adjustfigure{30pt}
\bibliographystyle{cambridgeauthordate}
\bibliography{jets}\label{refs}
 \printindex

\end{document}